\documentstyle[epsfig, 11pt]{article}

\def\reference{\parskip 0pt\par\noindent\hangindent 0.5 truecm}

\begin{document}
%
% Title
% Capitalise the title normally - do not use ALL CAPS.
%
\title{X-ray emission from GPS/CSS sources}
%

% Authors
% Here comes the author(s) of the paper. Please add the appropriate author
% names for your paper and indicate within the $^...$ the number(s)
% which corresponds to the institute(s) of each author. In this example
% the second author has two institutional affiliations.
% Add or remove authors as required.
% **** IMPORTANT: Leave the closing curly bracket line as is. ******

\author{Aneta Siemiginowska$^{1}$   
Thomas L. Aldcroft$^{1}$, Jill Bechtold$^{2}$, \\ Gianfranco
Brunetti$^{3}$, Martin Elvis$^{1}$, 
 Carlo Stanghellini$^{4}$ }

% IMPORTANT: leave this curly bracket as the first character of this line.

% Date - leave this blank.
\date{}
\maketitle

% Institutions
% Here fill in your institute name(s) and address(es)
% The number in $^...$ indicates the author number.  For example
{\center
$^1$ Harvard-Smithsonian Center for Astrophysics, 60 Garden St.,
Cambridge, MA 02138, USA, asiemiginowska@cfa.harvard.edu
$^2$ Steward Observatory, University of Arizona, USA,
$^3$ CNR, Bologna, Italy,
$^{4}$ CNR, Noto, Italy.
}

% Abstract
% Simply place your abstract between the \begin{abstract} and
% \end{abstract} commands.
%
\begin{abstract}
% Place the abstract here.
The high spatial resolution of the {\it Chandra} X-ray Observatory
allows us to study the environment of GPS/CSS sources to within an arcsec
of the strong compact core.  We present the discovery of X-ray jets
in two GPS quasars, PKS1127-145 and B2 0738+393, indicating that
X-ray emission associated with the relativistic plasma is present at
large distances from the GPS nucleus. We also discuss first results
from {\it Chandra} observations of our GPS/CSS sample. We find that 6
out of 10 sources show intrinsic absorption at a level which may be
sufficient to confine the GPS source.
\end{abstract}

{\bf Keywords: galaxies: active --- X-rays: galaxies}
% Place keywords here. Please write all keywords in lower case. PASA uses the
 %standard list of subject 
% headings adopted by The Astrophysical Journal and available from URL:
%   http://www.journals.uchicago.edu/ApJ/keywords_text.html

% A formatting command to add space between the author list and the body
% of the paper when printed. This spacing may be changed as desired.
\bigskip

%
% Body of paper
%

%\section{Previous X-ray Studies}

\section{Introduction}
% Place contents of first section here.

Giga-Hertz Peaked (GPS) and Compact Steep Spectrum (CSS) sources are
major classes of radio sources that may be offering glimpses of the
early stages of radio source formation. They could well offer
insight into the physical processes triggering activity in galactic
nuclei and so into the causes of quasar evolution. Although these
sources are now receiving considerable attention in the radio band,
they remain neglected at other wavelengths.

During the last two decades there has been {\em no} systematic X-ray
study of GPS/CSS sources and only a few X-ray observations have been
performed.  O'Dea (1998) lists 31 sources with available X-ray
information, including 7 sources with upper limits only. This sample
shows, as expected, that galaxies are less luminous in X-rays than
quasars.  Elvis et al (1994) observed X-ray absorption in two out of
three high redshift GPS quasars, suggesting that their environment
might be different from other quasars.  ROSAT upper limits for a few
GPS/CSS galaxies of L$_X < 3 \times 10^{42}$ ergs~s$^{-1}$ are
consistent with the X-ray emission expected from poor clusters or
early type galaxies.  O'Dea et al (2000) presented the first X-ray
detection of a GPS galaxy, which had L$_X \sim 2 \times 10^{42}$
ergs~s$^{-1}$ with ASCA.  The highest spatial resolution observations
before {\it Chandra} were made with the ROSAT HRI in which 2 out of 4
GPS quasars showed traces of an extended emission (Antonelli \& Fiore
1997).
	
The lack of available X-ray information is surprising, since strong
X-ray emission is expected to be associated with strong radio power
and could in principle help in understanding the emission processes
and dynamics related to radio-emitting structures.  Is the X-ray
emission directly connected to the expanding jet? X-ray emission
associated with a local halo or cluster (T$\sim$ 10$^7$K) could help
in studying the environment of the GPS/CSS sources. We should be able
to look for the remnants of a merger event or any signatures of
interactions between an expanding radio plasma and the IGM. X-ray
spectra can constrain the total absorbing column, so we may be able to
check whether there is sufficient material to confine the GPS source.

X-rays can also help to understand any link between GPS and large scale
radio sources and answer questions related to the source
evolution: Can we see any evidence for intermittent activity?  Do GPS
sources grow into large radio galaxies?  

The {\it Chandra} X-ray Observatory, with its exceptional image
quality (PSF FWHM of $\sim$0.5 arcsec), has discovered many X-ray jets
associated with radio structures in radio galaxies and quasars (see
http://hea-www.harvard.edu/XJET/). {\it Chandra} is the only X-ray
telescope which can resolve structures on 1 arcsec scales, and is
well-suited for studying the environment of GPS/CSS sources.  Are
X-ray jets present in GPS sources?  Can {\it Chandra} detect extended
emission hinted at with ROSAT HRI images for two GPS quasars?

Here, we present {\it Chandra} observations of two GPS sources,
PKS~1127-145 and B2~0738+313, in which we have discovered X-ray jets.
We also present preliminary analysis of our GPS/CSS sample observed
during the first half year of the {\it Chandra} AO3 cycle.
We assume H$_{0}$=50~km~sec$^{-1}$ Mpc$^{-1}$ and q$_{0}$=0.5.

\subsection{PKS~1127-145 and B2~0738+313}

We observed the two GPS quasars with the {\it Chandra} ACIS-S
detector.  PKS~1127-145 is a quasar at z=1.187 with $L_{bol} = 5\times
10^{46}$~ergs~sec$^{-1}$ and a GPS radio spectrum that peaks at
$\sim$1GHz. It also has an intervening damped Lyman $\alpha$ system
(DLA) at $z_{abs}= 0.312$ (Bechtold et al 2002). The {\it Chandra}
ACIS-S observation shows a $\sim 30$~arcsec X-ray jet; at the quasar
redshift it is $\sim 330 h_{50}^{-1}$~kpc projected on the sky
(Fig.1). As described in Siemiginowska et al (2002a), the jet curves
and its emission is very faint in comparison to the nucleus.  The
ratio of individual knots intensities to the core intensity is $\sim
1:450$.  The X-ray knots correspond to the VLA knots only roughly,
with X-ray peak intensities preceding the radio peak intensities. The
X-ray emission is stronger at the core, while the radio emission is
stronger away from the core.

B2~0738+313 is a quasar at z=0.63 and $L_{bol} \sim 10^{46}$
ergs~sec$^{-1}$. It has two DLA systems at z=0.0912 and z=0.2212 (Rao
\& Turnshek 1998).  The GPS radio spectrum peaks at $\sim 5$~GHz
(Stanghellini et al 1998).  ROSAT HRI observations show a hint of the
extended emission present on 10-20~arcsec scale. With a 27~ksec {\it
Chandra} ACIS-S observation we detected an X-ray jet extending up to
$\sim$35~arcsec away from the nucleus, with a few enhancements in the
form of hot spots and knots (Fig.1). The hot spot emission is faint
compared to the core, with intensity ratio of $\sim 1:200$. VLA radio
maps show faint extended radio emission in the form of lobes to the
north and south. The X-ray jet follows the radio emission to the south
(Fig.2). The X-ray jet emission is getting fainter moving away from
the core, while the radio emission has two hot spots at the end of the
southern lobe. There is NO X-ray emission corresponding to the
northern radio lobe with 3$\sigma$ upper limit of 2$\times 10^{-9}$
photons~cm$^{-2}$~sec$^{-1}$~pix$^{-1}$ (1 pix=0.164'').

\begin{figure*}
\begin{center}
\epsfig{figure=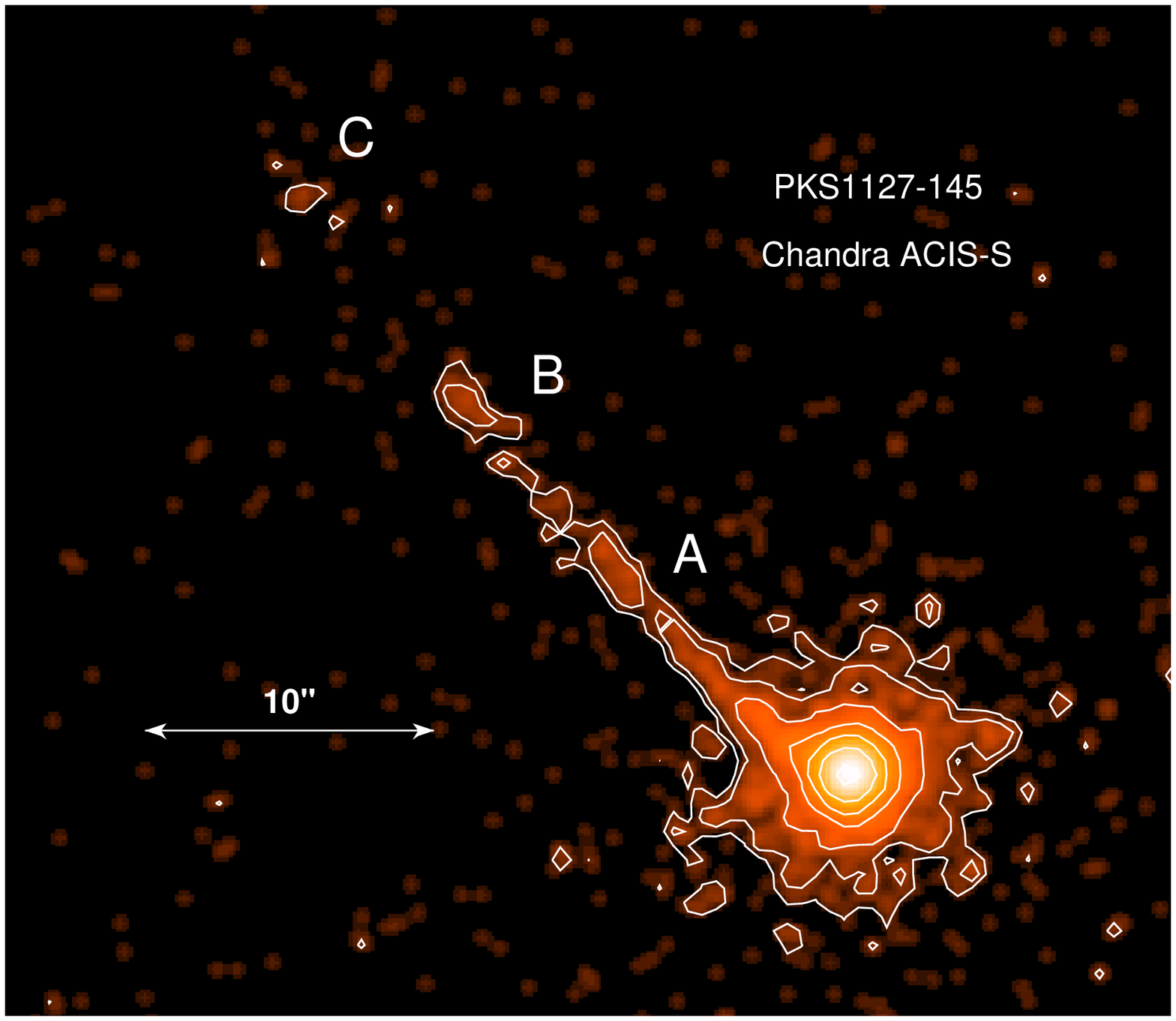,height=6.5cm,width=7.cm,angle=0}
\epsfig{figure=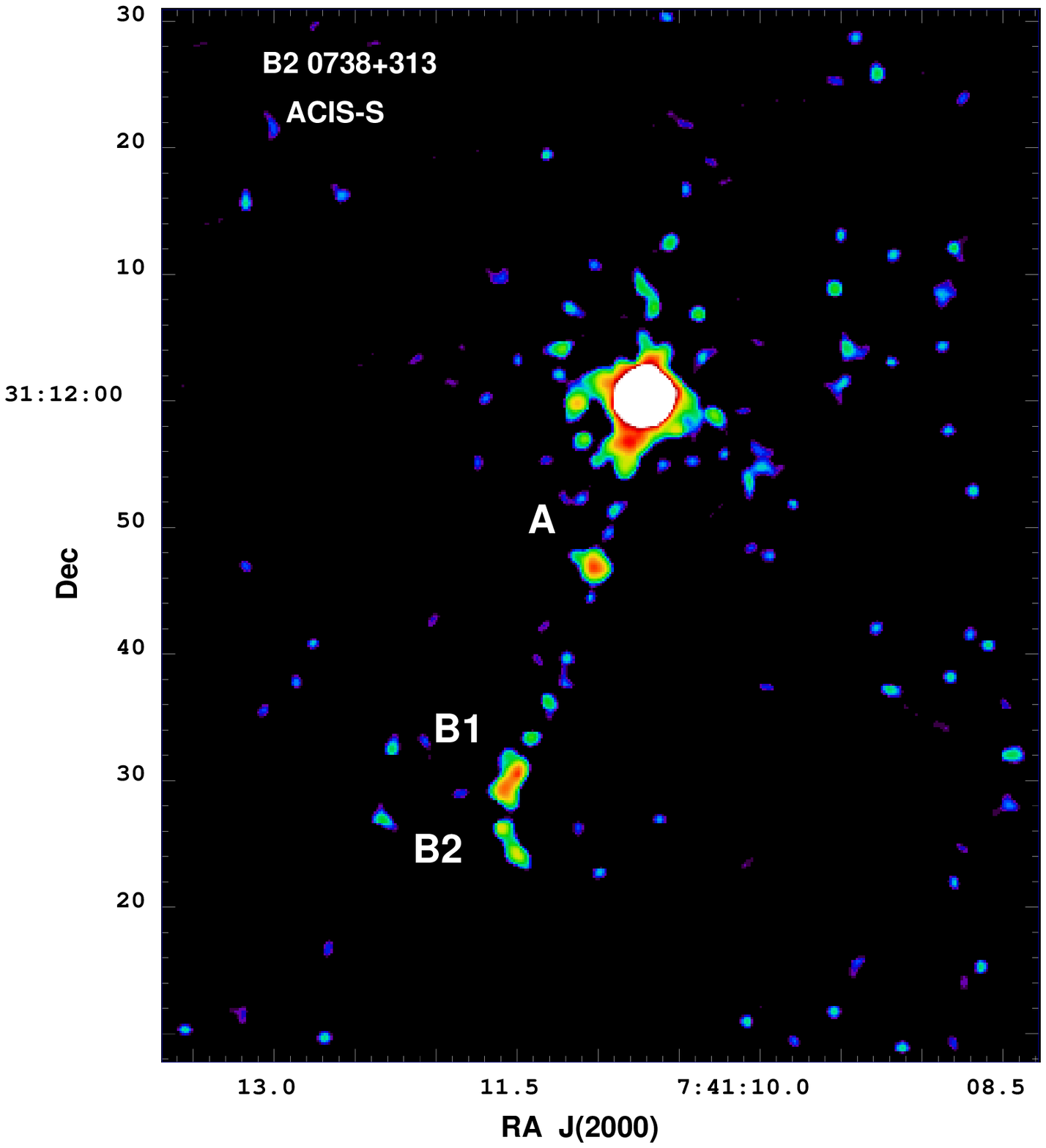,height=7.cm,width=7.5cm}
\end{center}
\vspace{-0.2in}
\caption{\small {\it Chandra} ACIS-S images smoothed 
with a Gaussian kernel (FWHM= 0.75~arcsec).  Only the photons with
energies between 0.3-6~keV have been included in the image. North is
up and east is left. (a) PKS~1127-145. The knots are labeled
A,B,C. Contour levels: 1.2,2.6,4.5,9,27,3000 counts/pixel
(Siemiginowska et al 2002a). (b) B2~0738+393.  The jet components,
A,B1 and B2 are indicated in the figure.  The peak emission at the
core corresponds to $4.5 \times
10^{-7}$~photons~cm$^{-2}$~sec$^{-1}$pixel$^{-1}$(1~pixel=0.164~arcsec)
(Siemiginowska et al 2002b).}
\label{fig1}
\end{figure*}

\subsection{Jet emission models}

Comparison of X-ray, optical and radio data rules out thermal
emission, synchrotron-self-compton (SSC) and simple direct synchrotron
emission (see Harris \& Krawczynski 2002 for review) as the origin of
the X-ray jet emission in PKS~1127-145. Inverse Compton scattering off
CMB photons (EIC/CMB) with moderate jet bulk velocities
($\Gamma_{bulk} \sim 2-3$) can readily accommodate the observations
because the EIC/CMB process is especially effective at high redshift
due to the $(1+z)^4$ scaling of the CMB. We note that X-rays from EIC
can trace the low energy ($\gamma \leq 10^3$) population of particles
which are not detectable in the radio band and thus delineate the
``fossil'' structure.

The X-ray emission in the two knots (B1,B2) of B2~0738+313 jet can be
explained by synchrotron emission. However, knot A needs an additional
component, because synchrotron emission requires unusually high
acceleration efficiencies. Note that in the synchrotron model the high
energy break needs to be at $\nu > 10^{19}$~Hz, which seems
unlikely. EIC/CMB process requires $\Gamma_{bulk}$ of $\sim$~10
(Fig.2) and might be a possible explanation for knot A X-ray emission.

\begin{figure*}
\begin{center}
\epsfig{figure=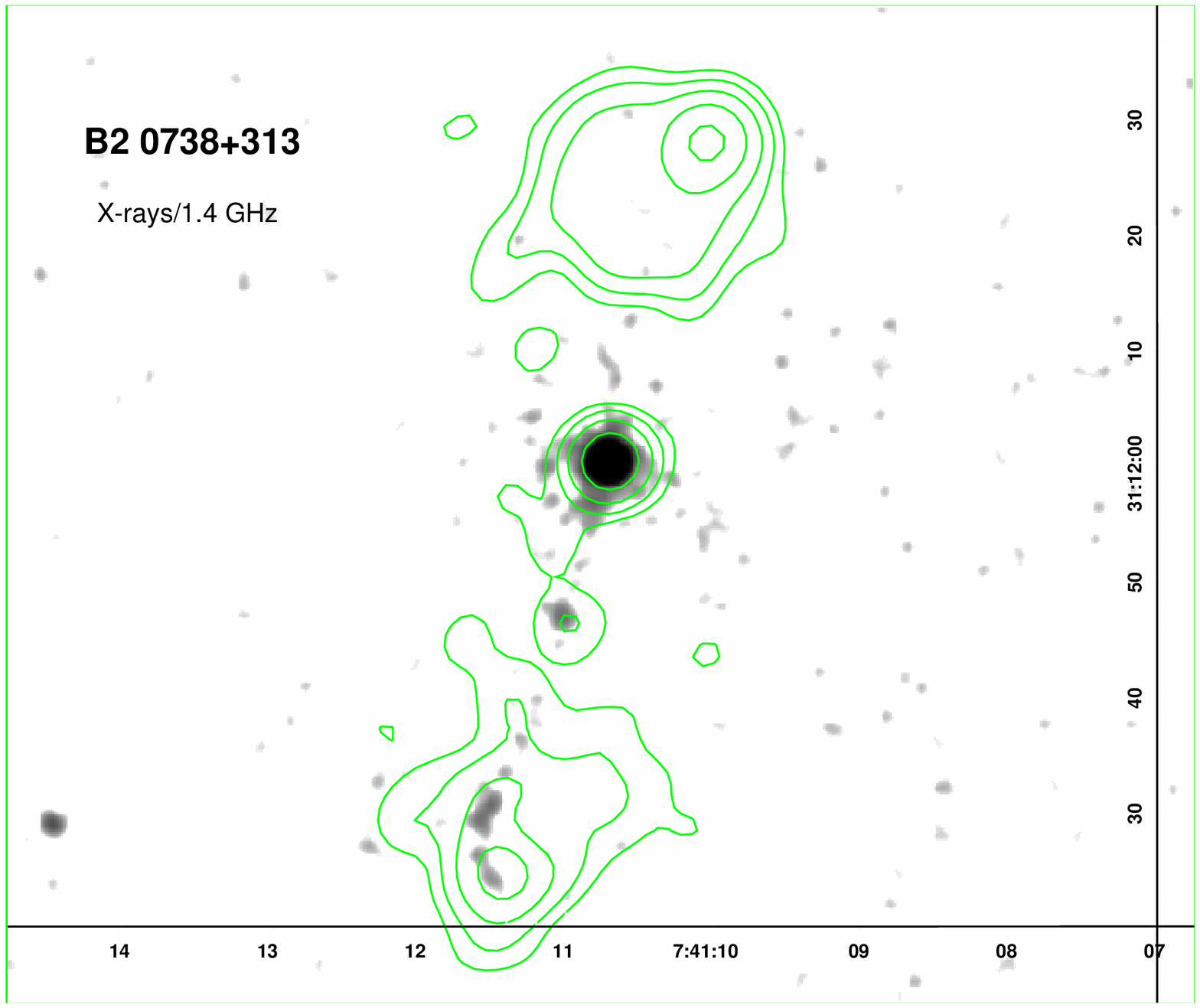, height=6.5cm,width=7.cm,angle=0}
\epsfig{figure=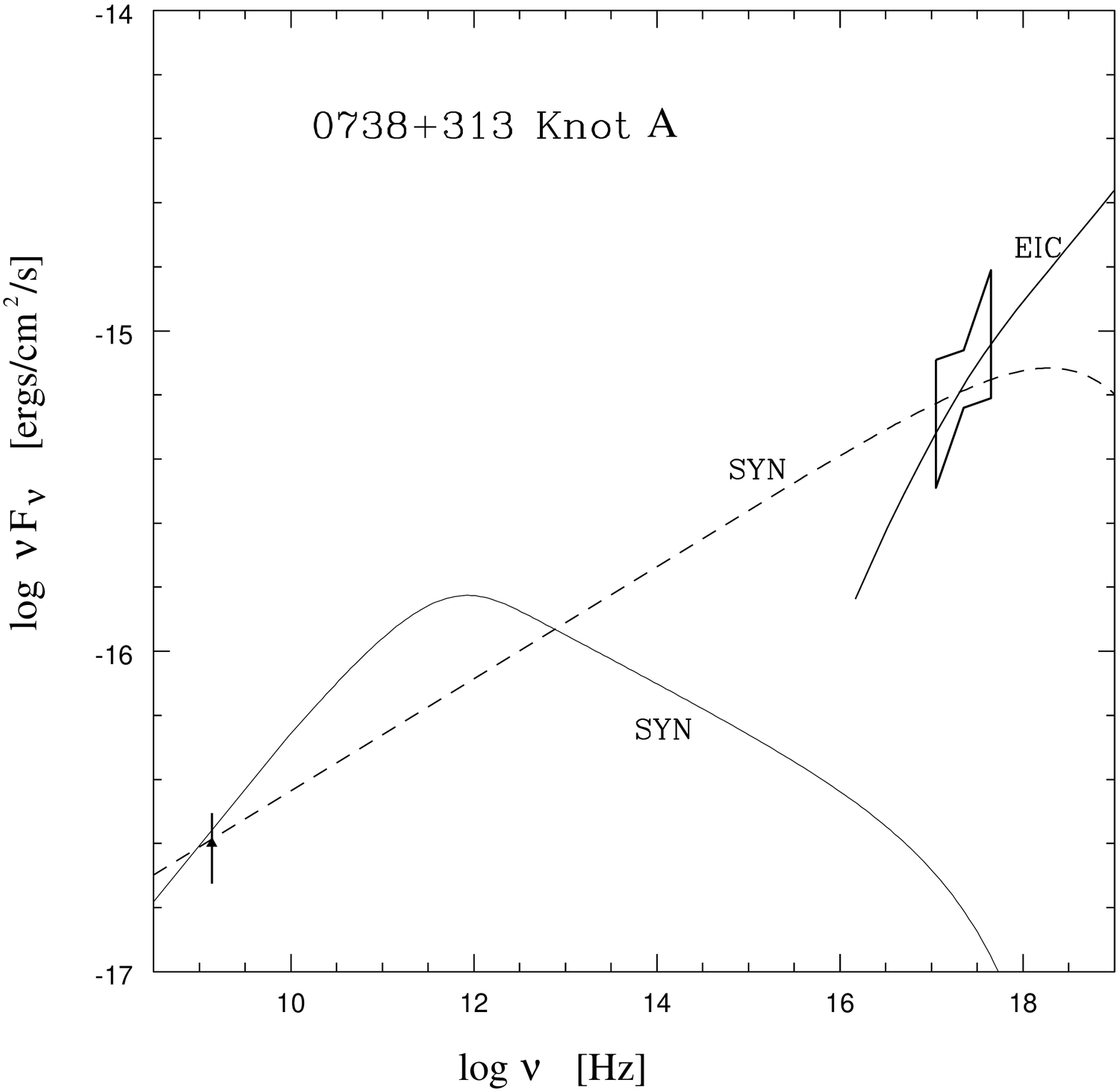,height=6.5cm,width=7.cm,angle=0}
\end{center}
\vspace{-0.3in}
\caption{\small (a) Superposition of B2~0738+313 {\it Chandra} X-ray image
(grey) with the radio (1.4 GHz) contours. The strong radio core has
been partly subtracted. Contour peak radio flux is at 1.47$\times
10^{-2}$ Jy/beam and contour levels are 2.5$\times
10^{-4}$(-3,3,6,12,25,50).  Grey scale in the X-ray image (0.3-6~keV)
varies from 10$^{-7}$ to 5$\times 10^{-7}$
photons~cm$^{-2}$~sec$^{-1}$~pixel$^{-1}$ (1 pixel=0.164
arcsec). North is up and east is left. (b) B2~0738+313: Jet emission
models for knot A. Dashed line show the synchrotron model with
required high energy cut-off of $>10^{19}$~Hz. Solid line shows
Inverse Compton model with CMB being a source of external photons
(EIC/CMB): $\alpha=$0.5-0.6, $\Gamma_{bulk}=10$, $\theta < 5$ deg,
B=15$\mu G$. Data points in radio and X-rays are shown.}
\label{fig2}
\end{figure*}

\subsection{Summary}

\begin{itemize}
\item X-ray jets in both sources extend far away from the nucleus,  
$\sim 300$~kpc projected distance.
\item The X-ray emission is most likely due to the interactions between
relativistic jet particles and Cosmic Microwave Background photons.
\item The X-ray jet emission indicates the presence of relativistic 
motion up to hundreds of kpc distance from the nucleus: $\Gamma_{\rm
bulk} \sim 3-10$.
\end{itemize}

\section{The first X-ray GPS/CSS Sample observed with {\it Chandra}}

The discovery of two X-ray jets associated with GPS quasars raises
more questions about the nature of the GPS sources: Are GPS/CSS
sources truly compact or do they have extended structures?  Are these
structures too faint to be easily detected in radio? Can we detect
extended emission in X-rays? Are GPS quasars simply extended radio
sources whose core is boosted towards us?

In order to address these questions we selected a sample for the first
systematic X-ray study of GPS/CSS sources using the following criteria:
(1) Radio size $ > 2 $ arcsec. 
(2) Redshift $ < 2 $.
(3) Low Galactic hydrogen column. 

We have obtained deep observations of the sources which had  
pre-existing measurements of their X-ray flux. For sources with no
previous X-ray observations we requested short 5 ksec exposures in
order to estimate their X-ray count rates and luminosity.

\begin{figure*}
\epsfig{figure=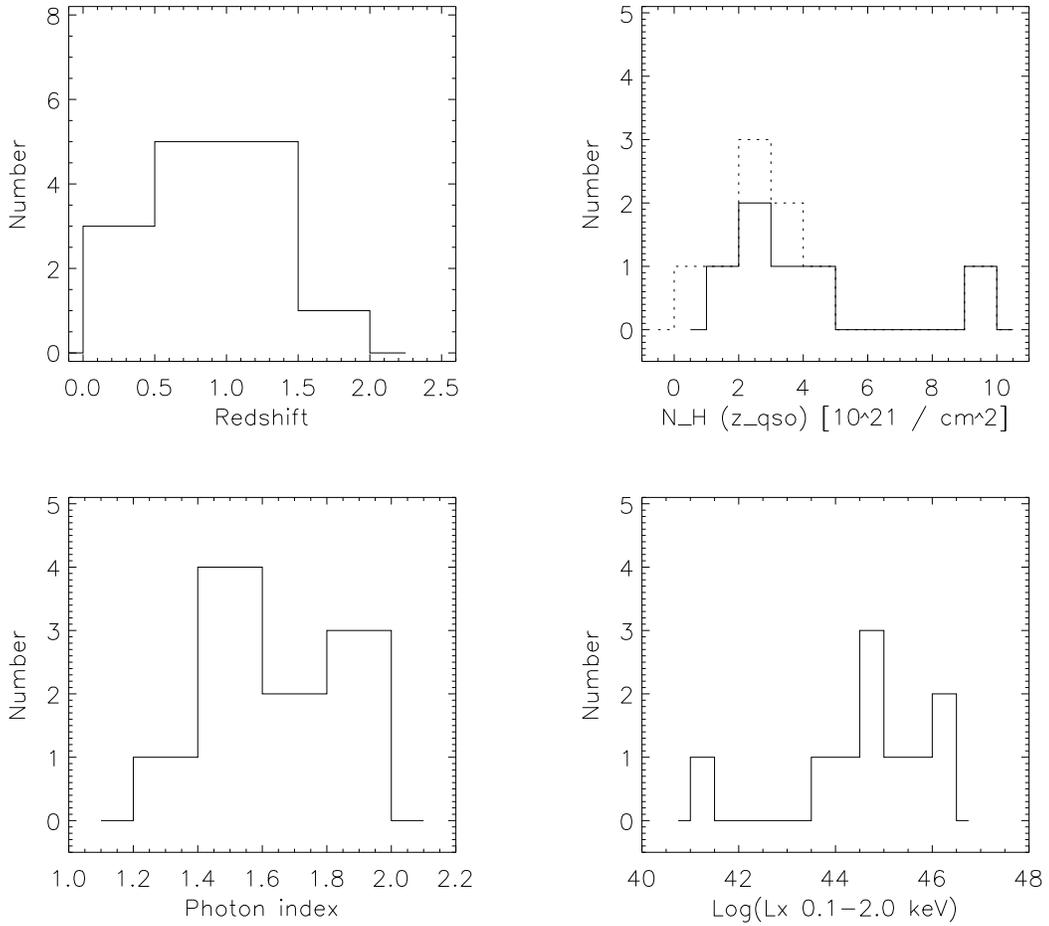, width=6in}
\vspace{-0.1in}
\caption{\small The {\it Chandra} GPS/CSS sample. (a) Redshift
distribution. (b) Distribution of the equivalent hydrogen column
density in excess of the Galactic column for the sample. Dashed line
indicates sources with the upper limits. (c) Distribution of photon
index within the sample. (d) Rest frame luminosity within 0.1-2~keV.
The GPS galaxy has the lowest X-ray luminosity in this plot.}
\end{figure*}
\label{fig3}
\subsection{First Results from {\it Chandra} Survey}

Thus far we have observed 10 out of 14 sources (Table 1). The sample
properties are presented in Fig.3. The details of data analysis
carried out with CIAO 2.2 software will be presented elsewhere
(Siemiginowska et al in preparation). Here we summarize our main
results.  All the observed sources have been detected and their
(0.1-2keV) fluxes range between:
$f_x \sim 0.1-2.8 \times 10^{-12}$erg~cm$^{-2}$~s$^{-1}$.
We found intrinsic absorption column in the excess of the Galactic
column in 6 sources: N$_H
\sim$10$^{21}-10^{22}$~cm$^{-2}$. 

We can obtain a lower limit on the density of the absorber assuming
that it provides the confinement within $\sim$10~kpc from the center.
This gives a density of 0.03-0.3~cm$^{-3}$, compared to the average
density of 1-10~cm$^{-3}$ required to permanently confine a jet in
numerical simulations of De Young (1993).

\begin{table}[h]
\begin{scriptsize}
\caption{GPS/CSS Chandra Sample}
\label{tab}
\begin{tabular}{lccccccccccl}
\hline
\hline\noalign{\smallskip}
Name	& redshift  &(N$_{H}$)$^a$ & (N$^{z_{abs}}_{H}$)$^b$ & $\Gamma ^{c}$ & f(0.1-2keV)$^{d}$ &
f(2-10keV) \\
&	 & E20 cm$^{-2}$ & E21 cm$^{-2}$&  & erg~cm$^{-2}$~s$^{-1}$ & erg~cm$^{-2}$~s$^{-1}$ \\	
\hline\noalign{\smallskip}
0941-080 & 0.228 &  3.67  & - & 1.5$\pm 2.4$ & 3.1e-15 & 7.1e-15\\
0134+329 & 0.367 & 4.54 & $<2.1$ & 1.96$\pm0.04$ & 1.82e-12 & 2.2e-12 \\
B2~0738+31& 0.63 & 4.18 & 2.1$\pm0.3$ & 1.56$\pm0.05$ & 2.98e-13(5.3) & 8.7e-13  \\
S5~0615+820&0.71 & 5.27 & $< 0.8$ & 1.61$\pm 0.59$& 1.22e-13 (1.22) & 2.8e-13 \\
1458+718 &0.905 &  2.33 & 1.2$\pm 0.2$ & 1.44$\pm 0.04$ & 8.02e-13 (10.8)& 2.24e-12 \\
1328+254 & 1.055&  1.08 & 2.2$\pm 0.2$ & 1.88$\pm 0.5$ &  2.54e-13 (3.9) & 4.01e-13 \\
1127-145 &1.187&  4.09 & 3.3$\pm0.2$ &1.24$\pm0.03$ & 1.25e-12(2.17) & 5.74e-12\\
1245-197 & 1.28  &  4.72  & $<3.6$ & 1.47$\pm0.02$ & 2.5e-14 & 6.6e-14 \\
1416+067 &1.439  &  2.5  & 4.5$ \pm0.3$ & 1.82$\pm0.03$& 1.45e-12 (2.8) & 2.88e-12 \\
1143-245 &1.95   &  5.22  & 9.4$\pm 5.2$ & 1.75$\pm0.39$ & 1.1e-13 (19.6) & 2.3e-13\\
\hline
%0740+380 &  5.64 &   \\
%1250+568 &  1.22 &   \\
%1829+29  & 11.16 && \\
%2128+048 &  5.23 && \\
\hline\noalign{\smallskip}
\end{tabular}

$^{a}$ N$_{H}$ - equivalent Galactic Hydrogen column density
$^{b}$ N$^{z_{abs}}_{H}$ - equivalent Hydrogen column density of the
absorber at the redshift of the source.
$^{c}$ $\Gamma$ - X-ray photon index  
$^{d}$ unabsorbed flux is shown in brackets.
Absorbed power law model : $A * E^{-\Gamma}*\rm exp^{- N_H
\sigma (E) - N^{z_{abs}}_H \sigma(E(1+z_{abs}))}$
~photons~cm$^{-2}$~sec$^{-1}$~keV$^{-1}$, where $A$ is the
normalization at 1~keV. $\sigma (E)$ and $\sigma E(1+z_{abs})$ are the
absorption cross sections (Morrison \& McCammon 1983, Wilms, Allen and
McCray 2000).

\end{scriptsize}
\end{table}

\section{Summary} 

We have discovered X-ray jets in two GPS quasars.  The jets travel up
to hundreds kpc from the active nucleus.  Our observations are
compatible with the intermittent activity model, in which old electrons
($\gamma
\sim 10^3$) from previous activity are detectable in X-rays, via
CMB Comptonization (LOFAR may detect directly), while new components
expand into the medium formed by the previous active phase.  However,
6 out of 10 GPS sources show intrinsic absorption columns: N$_H
>$10$^{21}$~cm$^{-2}$ $=>$ n$_e > 0.03-0.3$~cm$^{-3}$ suggesting that
the sources could be both confined and intermittent.

Alternatively GPS quasars may simply be extended radio sources whose
core is boosted towards us (see Stanghellini this proceedings). 

% Place contents of next section here.  %

% Add as many section titles/contents as required.  

% % If you have subsections then use the % \subsection{Subsection Title} % command and if you have subsubsections then use the 
% \subsubsection{Subsubsection Title} 
% command.  To use these commands, 

% first remove the % from the start of the line.  
% It is preferable to embed your figures in the text.  
% One way to do this is to use the psfig style file and use the following 

% commands to include the figures: 

% \begin{figure 
%\begin{center} 
% \psfig{file=filename.ps,height=10cm} 
% \caption{Write your figure caption here.}  
% \label{figlabel} % for cross-references
% \end{center}
% \end{figure} 

% To use the above commands, first remove the % from the beginning of 
% the lines and then fill in your own values etc as appropriate.  
% Tables % Please consult previous issues of PASA 
% to see how tables are to be formatted.
\section*{Acknowledgments} 

% Place acknowledgments here. Omit above \section command if there 
% are no acknowledgments.
This research is funded in part by NASA contracts NAS8-39073.  Partial
support for this work was provided by the National Aeronautics and
Space Administration through Chandra Award Number GO-01164X and
GO2-3148A issued by the Chandra X-Ray Observatory Center, which is
operated by the Smithsonian Astrophysical Observatory for and on
behalf of NASA under contract NAS8-39073.

\section*{References} 

% PASA uses the same conventions as ApJ for journal abbreviations. Sample 
% references are as follows.  
% Please follow the same format for your references.  

%\reference Author, A. B., Anotherauthor, C. D. \and Thirdauthor, E. F. 1990, 
% PASA, 7, 350 

% for a journal article, or % \reference Author, A.B. \and Anotherauthor, C. D. 1990, in This is a Book

%Title, ed. C. D. Editor, (City: Publisher Name), 43

% for a book.  

{\small

\reference Antonelli, L. A.; Fiore, F., 1997, Memorie della Societa Astronomia Italiana, Vol. 68, p.299
\reference Bechtold, J., Siemiginowska, A., Aldcroft, T.L, Elvis,
M., Dobrzycki, A., 2002, ApJ, 562, 133
\reference Elvis, M., Fiore,F., Wilkes, B., McDowell, J., Bechtold,
J., 1994, Ap.J. 422, 60
\reference De Young, David S., 1993, ApJ,402, 95 
\reference  Harris, D.~E.~\& Krawczynski, H.\ 2002, ApJ, 565, 244 
\reference Lane, W., Briggs, F.H. \& Smette A., 2000, ApJ, 532,146
\reference Morrison, R.~\& McCammon, D.\ 1983, ApJ, 270, 119
\reference O'Dea, C.P. 1998, PASP, 110, 493 
\reference O'Dea, C.P.; De Vries, W.H.; Worrall, D. M.;
Baum, S.A.; Koekemoer, A, 2000, AJ, 119, 478
\reference Rao, S.M. \& Turnshek, D.A. 1998, ApJ, 500, L115
\reference Siemiginowska, A.; Bechtold, J.; Aldcroft, T.L. et al ;
2002a, ApJ, 570, 543
\reference Siemiginowska A., Stanghellini, C. 
Brunetti G.,  Fiore F.,  Aldcroft, T.L, etal 2002b, ApJ submitted;
\reference Stanghellini C.,  O'Dea C.P., Dallacasa D., Baum S.A.,
Fanti R., Fanti C. 1998, Astron. Astrophys. Suppl. Ser. 131, 303-315
\reference Wilms, J.,  Allen, A., \& McCray, R.\ 2000, ApJ, 542, 914

 % Add as many references as required.
}
\end{document}